\documentclass[11pt]{article}%
\usepackage{amssymb}
\usepackage{amsfonts}
\usepackage{amsmath}
\usepackage[nohead]{geometry}
\usepackage[doublespacing]{setspace}
\usepackage[bottom]{footmisc}
\usepackage{indentfirst}
\usepackage{endnotes}
\usepackage{graphicx}%
\usepackage{rotating}
\usepackage{float}
\usepackage{color}
\usepackage{booktabs}
\usepackage{pdflscape}
\usepackage{float}
\usepackage[T1]{fontenc}
\usepackage[utf8]{inputenc}
\usepackage{fancyhdr}
\usepackage{natbib}
\usepackage{multirow}
\usepackage{caption}
\usepackage{subcaption}

\usepackage{titling}
\date{}
\predate{}
\postdate{}

\makeatletter
\def\@biblabel#1{\hspace*{-\labelsep}}
\makeatother
\nonstopmode

\begin{document}
\sloppy

\title{\textbf{A comment on `Testing Goodwin: growth cycles in ten OECD countries'}}
\author{Matheus R. Grasselli\thanks{Corresponding author: grasselli@math.mcmaster.ca, Department of Mathematics and Statistics , McMaster University.
The authors are grateful for comments and suggestions by David Harvie. This research received partial financial support from the Institute for New Economic Thinking (Grant INO13-00011) and the Natural Sciences and Engineering Research Council of Canada (Discovery Grants).} \ and
Aditya Maheshwari\thanks{Department of Statistics and Applied Probability, University of California, Santa Barbara.}}

\maketitle

\begin{abstract}
We revisit the results of \citet{Harvie2000} and show how correcting for a reporting mistake in some of the estimated parameter values leads to 
significantly different conclusions, including realistic parameter values for the Philips curve and estimated equilibrium employment rates exhibiting 
on average one tenth of the relative error of those obtained in \citet{Harvie2000}.
\end{abstract}



\section{Introduction}
\label{intro} 

\citet{Harvie2000} represents a milestone in the empirical literature on the Goodwin model. Early attempts 
by \citet{Atkinson1969} and \citet{Solow1990} to bring the model to data were restricted to the United States and did not provide formal econometric estimates, but rather informal comparisons between features predicted in the model and quantities observed in the data. \citet{Desai1984} provided a breakthrough by fully estimating 
the original model and some of its extensions using data from 1855 to 1965, but restricted to the United Kingdom. By its turn, \citet{Harvie2000} offered the 
first comprehensive multi-country econometric estimation of the Goodwin model using data from ten OECD countries from 1959 to 1994 and proposed a systematic way to test the 
performance of the model, namely by comparing the econometrically estimated equilibrium points with the corresponding empirical averages for wage share and employment rates. 

The results reported in \citet{Harvie2000} were unequivocally negative: the estimated equilibrium points lie outside the observed cycles for all ten countries; the estimated equilibrium wage share exceeds the empirical average with a relative error of at least 20\% for all ten countries and more than 100\% for Greece; the estimated equilibrium employment rate is systematically below the empirical average with an absolute error of at least 2\% for all ten countries, more than 10\% for the UK, and more than 30\% for Germany (making the employment rate for this country higher than 100\%); the estimate period for the cycles varies between one and 2.5 years, whereas the observed trajectories indicate a much longer time scale for the cycles. In addition, consistently with the results previously obtained by \citet{Desai1984}, \citet{Harvie2000} found that Goodwin's assumption of real wage bargaining (that is to say, perfect inflation expectations and absence of money illusion) should be rejected for all countries in the sample, whereas the assumption of constant capital-to-output ratio should be rejected for all ten countries except Italy and the UK. 

Since its publication, \citet{Harvie2000} has been widely cited as a benchmark for empirical tests of dynamic growth cycle models, primarily by researchers 
that take the negative results reported in it as motivation to explore alternatives to the Goodwin model. Regrettably, these results are not 
reproducible because of a reporting mistake in \citet{Harvie2000}. In this note, we explain the mistake and its consequences, provide the corresponding 
corrected value using the estimates in \citet{Harvie2000}, and point to recent research showing that the performance of the Goodwin model and some 
of its extensions is not nearly as bad as previously reported. 

\section{The mistake and its consequences}

We briefly recall the relevant equations for the Goodwin model in the Appendix. To obtain the estimates $\hat \gamma$ and $\hat \rho$ for the parameters of the linear Philips curve, Harvie uses the autoregressive distributed lag (ARDL) model presented in equation (17) of \citet{Harvie2000}. The estimation results for this equation are 
presented in Table A2.3 of \citet{Harvie2000}. The key reporting mistake\footnote{We thank David Harvie for informing us about this mistake through private communication.} in this paper is that all the quantities shown in the top five rows of this table are off by a factor of 100. For 
example, the Constant term for Australia should be $-0.5313$ but is reported as $-53.13$ instead. As a result, because of the way the long-run coefficients are calculated from the coefficients in the ARDL model, as explained in the footnote on page 356 of \citet{Harvie2000}, the estimates for $\hat \gamma$ and $\hat \rho$ are also wrong 
by a factor of 100. For example, these coefficients should be $0.6236$ and $0.6710$ for Australia, but are reported in Table A2.3 as $62.36$ and $67.10$ instead. As can be seen from \eqref{lambda_equi}, provided $\hat \rho>0$, the equilibrium employment rates obtained from the wrong values of $\hat \gamma$ and $\hat \rho$ are necessarily smaller than the rates that would be obtained from the correct values. This observation alone explains most of the downward bias exhibited by the estimates for equilibrium employment rates reported in \citet{Harvie2000} when compared to their corresponding empirical averages. 

There are two additional mistakes in Table A2.3. Our calculation for the coefficient $\hat\rho$ for Germany using the values provided in the table gives $92.44$ instead of $65.55$, leading to an equilibrium employment rate of $0.93$, instead of $1.30$ as reported in Table 2 of \citet{Harvie2000}. When further corrected for the factor-of-100 mistake, the equilibrium employment rate obtained for Germany is 0.96. Similarly, our calculation for the coefficient $\hat\gamma$ for the United States gives $-8.42$ instead of $8.42$, leading to an equilibrium employment rate of $1.06$, instead of ``not possible to calculate'' as reported in Table 2 of \citet{Harvie2000}. When further corrected for the factor-of-100 mistake, the equilibrium employment rate for the United States is $0.86$.

We summarize the results of the correct calculations in Table \ref{CorrectTable} below, which should be used as a replacement for Table 2 in \citet{Harvie2000}. Notice that the first three rows show no change when compared to \citet{Harvie2000}, as there is no correction to be made in the estimates $\hat\alpha$, $\hat\beta$ and $\hat\sigma$. The next four rows show show Harvie's estimates (which are incorrect) for $\hat\gamma$ and $\hat\rho$ and the corresponding correct values. As can be seen from \eqref{omega_equi}, these values are not needed in order to calculate the estimates for equilibrium wage shares, so our own calculations for $\hat u^*$ are identical to the results reported in \citet{Harvie2000}, with the exception of differences in rounding for Finland, Germany, and Norway. We also provide the equilibrium wage share for the United States, which could have been calculated by Harvie, since this is not affected by the negative value obtained for $\hat\rho$ for this country, as incorrectly stated in the footnote to Table 2 in \citet{Harvie2000}. The values for the empirical averages $\bar u$ are taken from Table 1 of \citet{Harvie2000}. Next in Table \ref{CorrectTable} below, we show the incorrectly calculated values for the equilibrium employment rate $\hat v^*$ and their corresponding correct values - including the value for the United State, which turns out to be possible to calculate - followed by the empirical averages $\bar v$ also taken from Table 1 of \citet{Harvie2000}. Finally we show Harvie's incorrect estimate of the length of business cycle $T$ for each country and their corresponding correct values. The correct estimates are roughly 10 times larger than the values reported in \citet{Harvie2000} and are consistent with previous estimates reported in \citet{Atkinson1969} and \citet{Solow1990}.

\begin{table}[H]
\centering
\resizebox{\textwidth}{!}{%
\begin{tabular}{|l|r|r|r|r|r|r|r|r|r|r|}
\hline
Variable            & Australia & Canada & Finland & France & Germany & Greece & Italy  & Norway & UK      & US     \\ \hline
$\hat\alpha$            & 0.0166    & 0.0160 & 0.0303  & 0.0364 & 0.0329  & 0.0401 & 0.0460 & 0.0262 & 0.0221  & 0.0111 \\ \hline
$\hat\beta$             & 0.0226    & 0.0259 & 0.0080  & 0.007627 & 0.004142  & 0.003568 & 0.004918 & 0.0134 & 0.003690  & 0.0206 \\ \hline
$\hat\sigma$            & 2.4994     & 1.5698   & 3.1396    & 1.7974   & 2.4941   & 3.0292   & 3.3527   & 3.6710   & 2.5694    & 1.7751   \\ \hline
$\hat\gamma_{Harvie}$            & 62.36      & 59.01   & 32.00    & 54.85   & 85.49    & 46.02   & 71.24   & 118.07   & 18.54  & 8.42   \\ \hline
$\hat\gamma_{correct}$            & 0.6236      & 0.5901   & 0.3200    & 0.5485   & 0.8549    & 0.4602   & 0.7124   & 1.1807   & 0.1854    & -0.0842   \\ \hline
$\hat\rho_{Harvie}$              & 67.10      & 65.32   & 36.57    & 62.01   & 65.55    & 53.48   & 81.97   & 122.43   & 21.90    & -7.92 \\ \hline
$\hat\rho_{correct}$              & 0.6710      & 0.6532   & 0.3657    & 0.6201   & 0.9244    & 0.5348   & 0.8197   & 1.2243   & 0.219    & -0.0792 \\ \hline
$\hat u^*_{Harvie}$  & 0.90& 0.93 & 0.89 & 0.92 & 0.90 & 0.87 & 0.83 & 0.86 & 0.93 &  \\ \hline
$\hat u^*_{correct}$ & 0.90& 0.93 & 0.88 & 0.92 & 0.91 & 0.87 & 0.83 & 0.85 & 0.93 &  0.94 \\ \hline
$\overline{u}$ & 0.6867 & 0.7126 & 0.7023  & 0.6689 & 0.6904 & 0.4272 & 0.5592 & 0.6971 & 0.7588 & 0.7432  \\ \hline
$\hat v^*_{Harvie}$  & 0.93    & 0.90 & 0.88  & 0.89 & 1.30 & 0.86 & 0.87 & 0.96 & 0.85  &        \\ \hline
$\hat v^*_{correct}$ & 0.95    & 0.93 & 0.95  & 0.95 & 0.96 & 0.94 & 0.92 & 0.99 & 0.96  &  0.86     \\ \hline
$\overline{v}$ & 0.949 & 0.928 & 0.953 & 0.949 & 0.963 & 0.947 & 0.928 & 0.978 & 0.950 & 0.941 \\ \hline

$T_{Harvie}$        & 1.32      & 1.06   & 2.09    & 1.18  & 1.13    & 1.73   & 1.49   & 1.20    & 2.42    &        \\ \hline
$T_{correct}$     & 13.07     & 10.46   & 20.05    & 11.48   & 11.05    & 16.6   & 14.51   & 11.85   & 22.88    &  \\ \hline
\end{tabular}
}
\caption{Corrected values for Table 2 of \citet{Harvie2000}.}
\label{CorrectTable}
\end{table}

As we see in Table \ref{CorrectTable}, simply correcting from the reporting mistake in Table A2.3 of \citet{Harvie2000} leads to significant improvements in 
performance for the Goodwin model. With the exception of the anomalous case of the United States (for which both parameters in the Philips curve have the opposite sign as obtained for all other countries), the correct estimates for the equilibrium employment rates are only about 1\% away from the empirical averages. Another way of seeing this improvement is by observing that the average relative error in these estimates gets reduced tenfold from 
9.09\% for the values reported in \citet{Harvie2000} to a mere 0.60\% for the correct values above, excluding the United States, for which the equilibrium employment rate was not reported in \citet{Harvie2000}. If we include the anomalous case of the United States, the average relative error in employment rates is still just 1.40\%. 

\section{Further improvements}

Despite the marked improvements with respect to equilibrium employment rates observed in the previous Section, it remains the case that the results of \citet{Harvie2000}
show a poor agreement between the estimated equilibrium wage shares for the Goodwin model and their corresponding empirical averages. To address this problem one
needs to revisit the hypotheses behind the derivation of \eqref{goodwin_model_2}, including the choice of production function for the model. For example, as 
shown in \citet{GrasselliMaheshwari2016}, using a CES production function in the Goodwin model as proposed in \citet{vanderPloeg1985} leads to equilibrium estimates 
for wage shares that are much closer to empirical averages than those obtained for the original Goodwin model. 

\section{Conclusion}

We have shown how correcting a reporting mistake in \citet{Harvie2000} leads to significant improvements in the empirical performance of the Goodwin model. Apart from the quantitative changes noted in the previous section, this correction has qualitative implications, as several papers took the results of 
\citet{Harvie2000} as motivation for exploring methodological questions related to the Goodwin model. For example, \citet{VenezianiMohun2006} attributes the 
``puzzling econometric results obtained by Harvie (2000)'', including the ``rather unrealistic values of the Phillips curve parameters in all countries'' to the possibility of structural change in the model parameters. Taking our correction into account, however, eliminates the puzzles without having to resort to this explanation. Similarly, \citet{ColacchioSparroTebaldi2007} investigate the appearance of chaos in an extended version of the Goodwin model and rely on the estimates of \citet{Harvie2000} for the parameter values used in their numerical experiments. However, when the values of $\rho$ and $\gamma$ implied by our corrections are used, the point of bifurcation and time period of cycles turns out to be very different from what is reported in the study.

Above all, our correction aims to restore the status of the Goodwin model as a respectable starting point for more sophisticated dynamic models for growth cycles.

\begin{appendix}

\section{The Goodwin model}

Using the notation in \citet{Harvie2000}, the Goodwin model, first proposed in \citet{Goodwin1967}, consists of the differential equations 
\begin{align}
\label{goodwin_model_1}
\frac{\dot{u}}{u} &=-(\alpha+\gamma) + \rho v \\
\frac{\dot{v}}{v} &=\frac{1-\omega}{\sigma}-(\alpha+\beta)
\label{goodwin_model_2}
\end{align} 
for the wage share $u = w\ell/q$ and employment rate $v = \ell/n$, where $w\ell$ is the total real wage bill, $q$ is total real income, $\ell$ is the number of employed workers and $n$ is the total labour force. The constants $\gamma$ and $\rho$ arise from a linear Philips curve relating the change in real wage rate $w$ and 
the employment rate $v$:
\begin{equation}
\frac{\dot w}{w}=-\gamma+\rho v,
\end{equation}
whereas $\alpha$ and $\beta$ are constant growth rates in productive and labour force, and $\sigma$ is a constant capital-to-output ratio. The solution of this system of differential equations is a closed cycle around the equilibrium point 
\begin{align}
\label{omega_equi}
u^* & = 1-(\alpha+\beta)\sigma \\
v^* & = \frac{\alpha+\gamma}{\rho}
\label{lambda_equi}
\end{align}
with period given by 
\begin{equation}
\label{period}
T=\frac{2\pi}{\left[(\alpha+\gamma)(1/\sigma-(\alpha+\beta))\right]^{1/2}}.
\end{equation}
The test of the Goodwin model proposed by \citet{Harvie2000} consists of comparing the {\em econometric-estimate predictors} $(\hat u^*, \hat v^*)$ for the equilibrium point, which can be obtained from \eqref{lambda_equi}-\eqref{omega_equi} by substituting the econometric estimates for the underlying parameters in the model, with the empirical average of the observed employment rates and wage shares through the data sample. 

\end{appendix}

\bibliographystyle{apalike}
\bibliography{finance}

\end{document}